\documentclass[prl,twocolumn,amsmath,amssymb,superscriptaddress,showpacs]{revtex4}

\usepackage{graphicx}
\usepackage{epsfig}
\usepackage{epstopdf}

\begin{document}

\title{Electron Transport in Disordered Graphene Nanoribbons}

\author{Melinda Y. Han}
\affiliation{Department of Applied Physics and Applied
Mathematics, Columbia University New York, NY 10027}

\author{Juliana C. Brant}
\affiliation{Department of Physics, Columbia University New York,
NY 10027} \affiliation{Department of Physics Federal University of
Minas Gerais, Belo Horizonte, MG, Brazil}

\author{Philip Kim}
\affiliation{Department of Applied Physics and Applied
Mathematics, Columbia University New York, NY 10027}
\affiliation{Department of Physics, Columbia University New York,
NY 10027}

\date{\today}

\begin{abstract}
We report an electron transport study of lithographically
fabricated graphene nanoribbons of various widths and lengths at
different temperatures. At the charge neutrality point, a
length-independent transport gap forms whose size is inversely
proportional to the width. In this gap, electron is localized, and
charge transport exhibits a transition between simple thermally
activated behavior at higher temperatures and a variable range
hopping at lower temperatures. By varying the geometric
capacitance through the addition of top gates, we find that
charging effects constitute a significant portion of the
activation energy.
\end{abstract}

 \maketitle

In recent years graphene has been celebrated for its potential as
a new electronic material~\cite{Geim07, Geim08}. However, the
absence of an energy band gap in graphene poses a challenge for
conventional semiconductor device operations. Previous work
\cite{Han07, Chen07, Li-Dai08} has shown that this hurdle can be
overcome by patterning graphene into nanometer size ribbons or
constrictions. The resulting transport gap formation can be most
simply attributed to quasi 1-dimensional (1D) confinement of the
carriers, which induces an energy gap in the single particle
spectrum~\cite{ManyGNRs}. Detailed experimental studies of
disordered GNRs~\cite{Ponomarenko08, Stampfer09, Molitor09,
Todd09, Liu08, Gallagher09},  , however, suggest that this
observed transport gap may not be a band gap.  In an effort to
explain these experimental results, various theoretical
explanations for the transport gap formation in disordered
graphene nanostructures have been proposed, including models based
on Coulomb blockade in a series of quantum dots~\cite{ Sols07},
Anderson localization due to edge
disorder~\cite{Gunlycke-White07APL, Lherbier08, Evaldsson08,
Querlioz08, Mucciolo09, Martin09}, and a percolation driven
metal-insulator transition~\cite{Adam08}. Further systematic
experiment is necessary to distinguish between these different
scenarios.

In this letter, we study the scaling of the transport gap in
graphene nanoribbons (GNRs) with various widths and lengths at
different temperatures. We find that four different energy scales
can be extracted from transport measurement. From the scaling of
these characteristic energies with GNR width and length, we find
evidence of a transport mechanism in disordered GNRs based on
hopping through localized states whose size is close to the width
of the GNRs.

GNRs with different lengths ($L$) and a widths ($W$) were
fabricated following the procedures described in~\cite{Han07}.
Most experiments in this report were performed on back-gated GNRs
on a substrate of highly doped silicon with a 285 nm thick SiO$_2$
gate dielectric. An example of such a device is shown in the inset
to Fig.~1(a). We measured electron transport in a total of 41 GNRs
with $20<W<120$~nm and $0.5<L<2\mu$m at different temperatures
$1.5<T<300$~K. Additionally, we fabricated top-gated GNRs with
15~nm of hydrogen silsesquioxane (HSQ) and 10~nm of HfO$_2$ as the
gate dielectric material. The increased capacitive coupling
allowed a comparative study of charging effects in back gated
GNRs.

\begin{figure}
\includegraphics[width=1.0\linewidth]{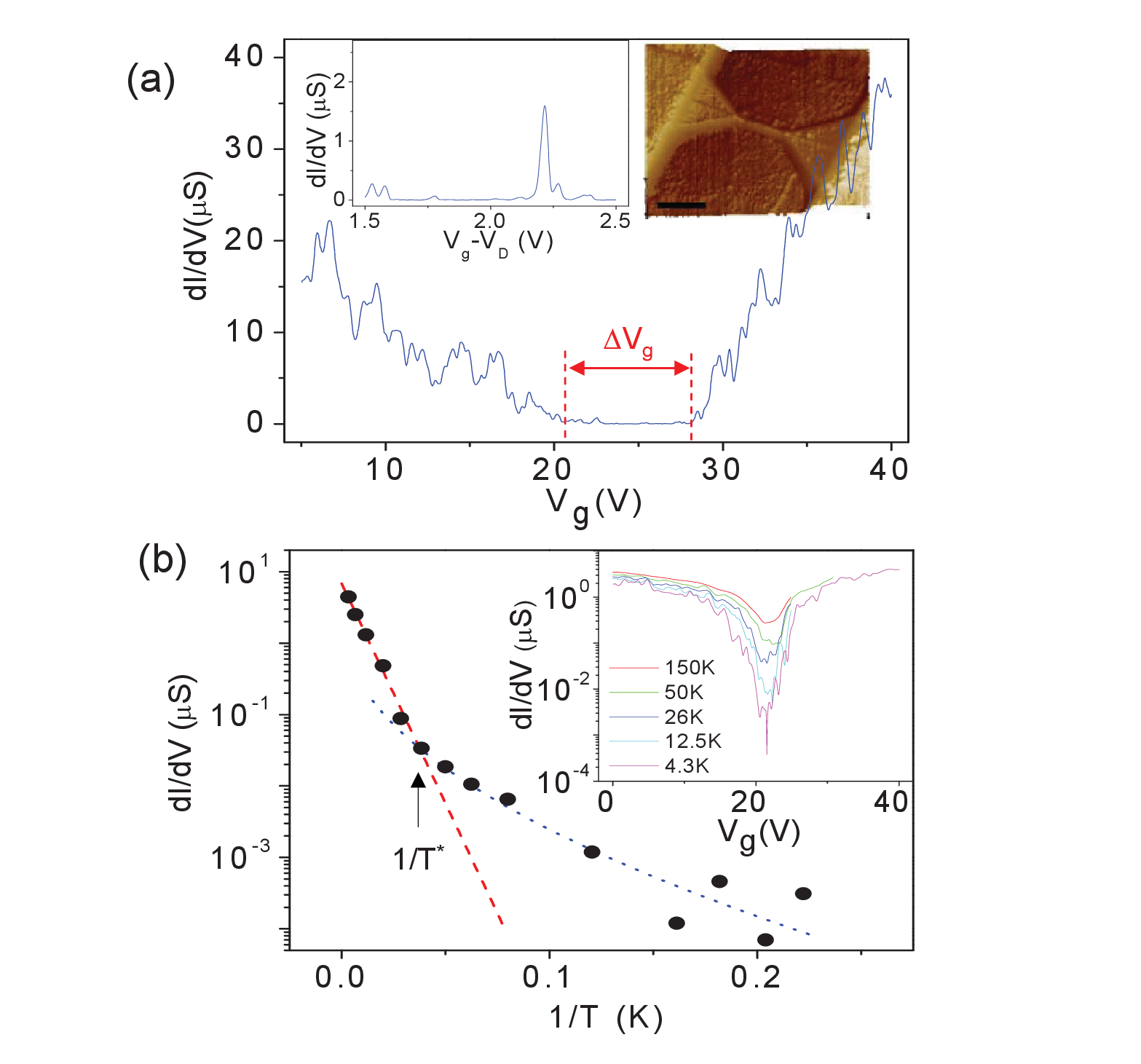}
\caption{ (a) Differential conductance of a GNR with $W=36$~nm and
$L=500$~nm, plotted as a function of back gate voltage. Dashed
lines highlight measurement of $\Delta V_g$.  Right inset shows an
atomic force microscope image of the device. Scale bar is 500~nm
Left inset shows a close-up of conductance within the gap regime
plotted as a function of $V_g-V_D$, where $V_D=21$~V is the gate
voltage for the charge neutrality point. (b) $T$ dependence of the
minimum conductance of the same GNR in (a). The dashed line is a
fit to simple activated behavior; the dotted line is a fit to
variable range hopping with $\gamma =1/2$ and $T_0=460$~K. An
arrow highlights the position of $T^*$. Inset shows conductance
versus $V_{g}$ at several temperatures.} \label{Fig.1}
\end{figure}

GNR conductance is strongly suppressed for a region of back gate
voltages $V_g$ near the graphene charge neutrality
point~\cite{Han07, Chen07, Stampfer09, Molitor09, Todd09, Liu08,
Gallagher09}, suggesting the formation of a transport gap.
Fig.~1(a) shows low bias differential conductance $G=dI/dV$ as a
function of $V_g$ for a typical GNR. The transport gap region in
back gate voltage, $\Delta V_{g}$, can be identified in this curve
by extrapolating the smoothed $dG/dV_g$ to zero~\cite{Stampfer09,
Gallagher09}. We note that reproducible conductance peaks appear
in the gap region~\cite{Stampfer09, Molitor09, Gallagher09}  (left
inset Fig.~1(a)), which are indicative of resonant conduction
paths through localized states inside the transport gap. In
general, resonance peaks in the gap are less than 10~\% of the $G$
values outside of the gap region.

The observed transport gap, $\Delta V_{g}$ corresponds to an
energy in the single particle energy spectrum: $\Delta _{m} =\hbar
v_F \sqrt{2\pi C_g \Delta V_{g}/|e|}$, where $v_F=10^6$~m/sec is
the Fermi velocity of graphene~\cite{CastroNeto09} and $C_g$ is
the capacitive coupling of the GNR to the back gate. This
geometric capacitance is strongly dependent on ribbon dimensions
and we calculate it using a finite element model, obtaining, for
example, $C_g=690$~aF/$\mu$m$^2$ and $\Delta_{m}=200$~meV for the
particular device in Fig 1.

Away from the small resonant conductance peaks, the conductance is
strongly suppressed in the transport gap, and the dominant charge
transport can be described by thermally excited hopping between
localized states~\cite{footnote1}. We study the thermal activation
of the off-resonant conduction in this regime by measuring
$G_{min}$, the minimum conductance for a given sweep of gate
voltage $V_g$, at different temperatures (inset to Fig.~1(b)).
Fig.~1(b) shows an Arrhenius plot for $G_{min}(T)$. Evidently,
thermally excited transport exhibits two distinct behaviors at low
and high temperature regimes, respectively, separated by a
characteristic temperature $T^*$. At high temperatures ($T>T^*$),
the transport is simply activated: $ G_{min}\sim
\exp(-E_a/2k_BT)$, where $E_a=285$~K is obtained from a linear fit
of the Arrehenius plot (dashed line). At lower temperatures
($T<T^*$), however, $G_{min}$ deviates from the simple activation
behavior and decreases more slowly with decreasing temperature
than the activated transport would imply. In this low temperature
regime, the overall behavior is consistent with variable range
hopping (VRH), where $G\sim \exp(-(T_0/T)^{\gamma})$, with $1/3
\leq \gamma \leq 1/2$ and a constant $T_0$, determined by the
characteristics of the localized states~\cite{VRHref1}.

\begin{figure}
\includegraphics[width=1.0\linewidth]{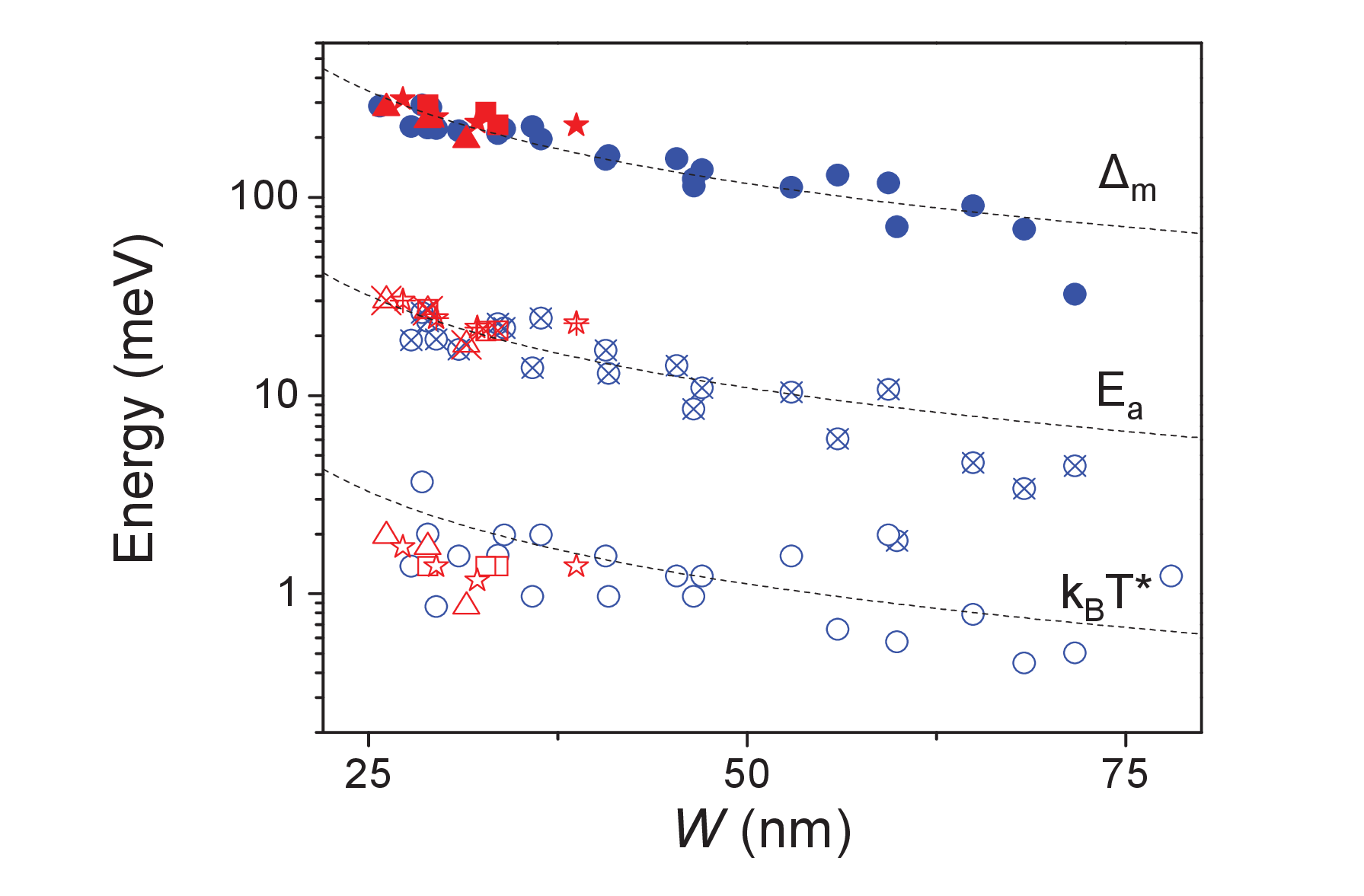}
\caption{ GNR transport energy scales : $\Delta_m$ (solid), $E_a$
(shaded), and $k_BT^*$ (open) plotted as a function of GNR width.
Circles correspond to ribbons of $L=500$~nm.  Triangles, squares,
and stars correspond to ribbons of length 1, 1.5, and 2~$\mu$m,
respectively. The dashed lines are the fits described in the
text.} \label{Fig.2}
\end{figure}

The aforementioned GNR transport gap and temperature dependent
characteristics are typical of all GNRs with $W\lesssim 80$~nm, so
that $\Delta_m$, $E_a$, and $k_BT^*$ can be determined for each of
these narrow GNRs.  These three representative energy scales are
plotted as a function of $W$ in Fig 2. In this graph, we note that
(i) there is a clear separation between these energy scales,
setting a general relation: $\Delta_m>E_a>k_BT^*$ for given $W$;
(ii) $\Delta_m$, $E_a$, and $T^*$ depend sensitively on $W$ but
not $L$; and (iii) the energy scales are reasonably well described
by inverse proportion to the lateral confinement of the GNR.  The
length independence can be noticed by comparing characteristic
energies of the GNRs with similar $W$ but different $L$
(represented by different symbols in Fig.~2), and suggests that
these three energy scales are 1D intensive properties of GNRs.  To
show this, we define the normalized width $w=(W-W_0)/a_0$, where
$a_0=0.142$~nm is the carbon-carbon bond length and $W_0$ is an
offset introduced phenomenologically. Then, we find that all
energy scales can be reasonably fit (dotted lines):
$\Delta_m=\Delta_m^0/w$; $E_a=E_a^0/w$; $T^*=T^*_0/w$ with the
proportionality parameters $\Delta_m^0=36.3$~eV, $E_a^0=3.39$~eV,
and $k_BT^*_0=347$~meV, respectively, with $W_0=12$~nm held fixed
for all three fits~\cite{footnote2}.

Edge disorder in the GNRs tends to induce wavefunction
localization, with a localization length that decreases rapidly
with decreasing energy, resulting in a transport gap with strongly
localized states at energies between the mobility
edges~\cite{Gunlycke-White07APL}. The size of this mobility gap is
larger than the clean band gap of an ideal ribbon; Querlioz {\it
et. al.} calculate the scaling prefactor $\Delta_m^0\approx
32.2$~eV, averaged over many configurations of edge
disorder~\cite{Querlioz08}. The close match of our data to
theoretical prediction supports the view that atomic defects at
the graphene edges create localized states. We point out, however,
that the observed energy scales lie within the range of disorder
potential fluctuation created by the charged impurities in the
SiO$_2$ substrate~\cite{Adam08}, making it difficult to exclude
the contribution of a substrate disorder induced transport gap, as
discussed in a recent experiment on transport in thermally
annealed GNRs~\cite{Gallagher09}.

On the other hand, $E_a^0/\Delta_m^0\approx 0.1$; i.e., the
activation energy at higher temperatures is an order of magnitude
smaller than $\Delta_m$. This observation excludes the scenario
that extended states carry current via thermal activation across
the transport gap. Instead, we interpret the simply activated
behavior as a signature of 1D nearest neighbor hopping (NNH)
through localized states within the transport gap~\cite{Martin09}.
In this picture, disorder at the edges tends to produce a rapid
variation in the local density of states over the whole width of
the ribbon, blocking the conductive paths and leading to a
quasi-1D arrangement of localized states~\cite{Evaldsson08}.
Martin and Blanter predict~\cite{Martin09} that the energy spacing
between nearest neighbor states is determined by $\sim t'/w$,
where $t'\approx 0.2 t$ is the hopping matrix element between
second nearest neighbor carbon atoms in graphene, so that
$E_a^0\sim 2 t'=1.2$~eV~\cite{Martin09}.  Our measured value for
this scaling prefactor, 3.39~eV, is somewhat larger than this
prediction, which may be explained by the contribution of a
charging energy to the hopping energy $E_a$, discussed in more
detail below.

The change of the transport behavior across the temperature $T^*$
allows a further comparison of our data to theory. In a very
recent theoretical work, the NNH and VRH crossover is calculated
to occur at $T^* = E_a/k_B\alpha$, where $\alpha\approx 8$ was
estimated numerically~\cite{FoglerPreprint}. In our experiment, we
obtain $E_a^0/k_B T^*_0=9.8$, reasonably consistent with this
theoretical prediction, lending further support to a model of
charge transport via thermally activated hopping between localized
states.

\begin{figure}
\includegraphics[width=1.0\linewidth]{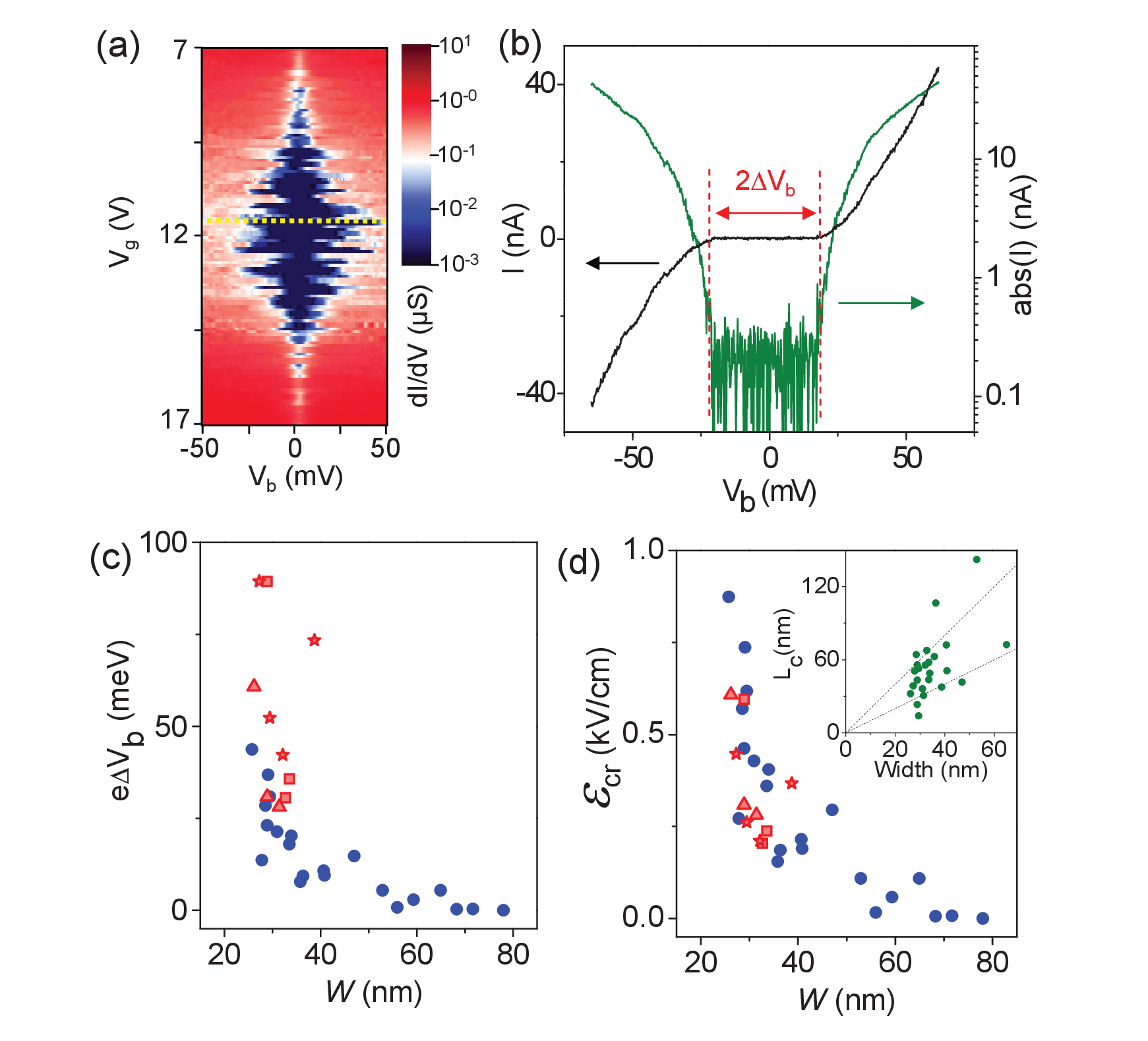}
\caption{(a) Differential conductance as a function of $V_g$ and
$V_b$ measured in a GNR with $L=1$~$\mu$m and $W=31$~nm. (b)
Current as a function of $V_b$ with $V_g$ fixed in the
off-resonant condition marked by the dotted line in (a). $\Delta
V_b$ is highlighted by the vertical dashed lines. (c) $\Delta V_b$
as a function of $W$. Symbol shapes in (c) and (d) represent
different GNR lengths following the convention set in Fig.~2 (d)
The critical electric field $\mathcal{E}_{cr}$ versus $W$
converted from the data set in (c).} \label{Fig.3}
\end{figure}

An alternative approach to probing the GNR transport gap is
measurement of the non-linear transport
characteristics~\cite{Han07}. Fig.~3(a) shows differential
conductance, $dI/dV_b$ as a function of $V_g$ and source-drain
bias voltage $V_b$. Transport through the GNR at finite $V_b$
shows a strong non-linear $I-V_b$ characteristic when $E_F$ is in
the transport gap regime, which is most extreme when $V_g$ is near
the charge neutrality point of the GNR (Fig.~3(b), black curve).
The non-linear gap $\Delta V_b$ can be defined where a steep
increase of current appears in logarithmic scale (Fig.~3(b), green
curve).

In our previous study~\cite{Han07}, the energy corresponding to
$e\Delta V_b$ was interpreted to be the band gap of the GNR.
However, this naive interpretation should be carefully
reconsidered for edge disordered GNRs, where the charge transport
is dominated by hopping through localized states. Indeed, from the
plot of $\Delta V_b$ vs $W$ (Fig 3(c)), we notice that $\Delta
V_b$ depends strongly on $L$, and is not well determined by $W$
alone, unlike the previous three characteristic energy scales
($\Delta_m$, $E_a$, and $k_B T^*$). Since the charge transport in
the disordered GNRs is diffusive, it is likely that electric field
is driving transport in the transport gap. Indeed, if we convert
$\Delta V_b$ into the corresponding critical electric field
$\mathcal{E}_{cr}=\Delta V_b/L$, we restore a reasonable scaling
behavior, where $\mathcal{E}_{cr}$ depends only on $W$ and not on
$L$ (Fig.~3(d)).

In disordered systems in which transport is dominated by hopping
through localized states, applied electric field $\mathcal{E}$
plays a similar role to temperature. Thus we can treat the
electric field as an effective temperature:
$k_BT_{eff}=e\mathcal{E}L_c$, where $L_c$ is the averaging hoping
length between localized states~\cite{Shklovskii1973hci}. Noting
that the transition from NNH dominated transport to VRH transport
occurs at $T^*$, we can estimate $L_c \approx k_B
T^*/e\mathcal{E}_{cr}$. For most GNRs in this experiment we find
that $W \lesssim L_c < 2W$ (Fig.~3(d) inset). The fact that $L_c
\gtrsim W$ supports our claim that hopping transport through the
ribbons is effectively 1D.

\begin{figure}
\includegraphics[width=1.0\linewidth]{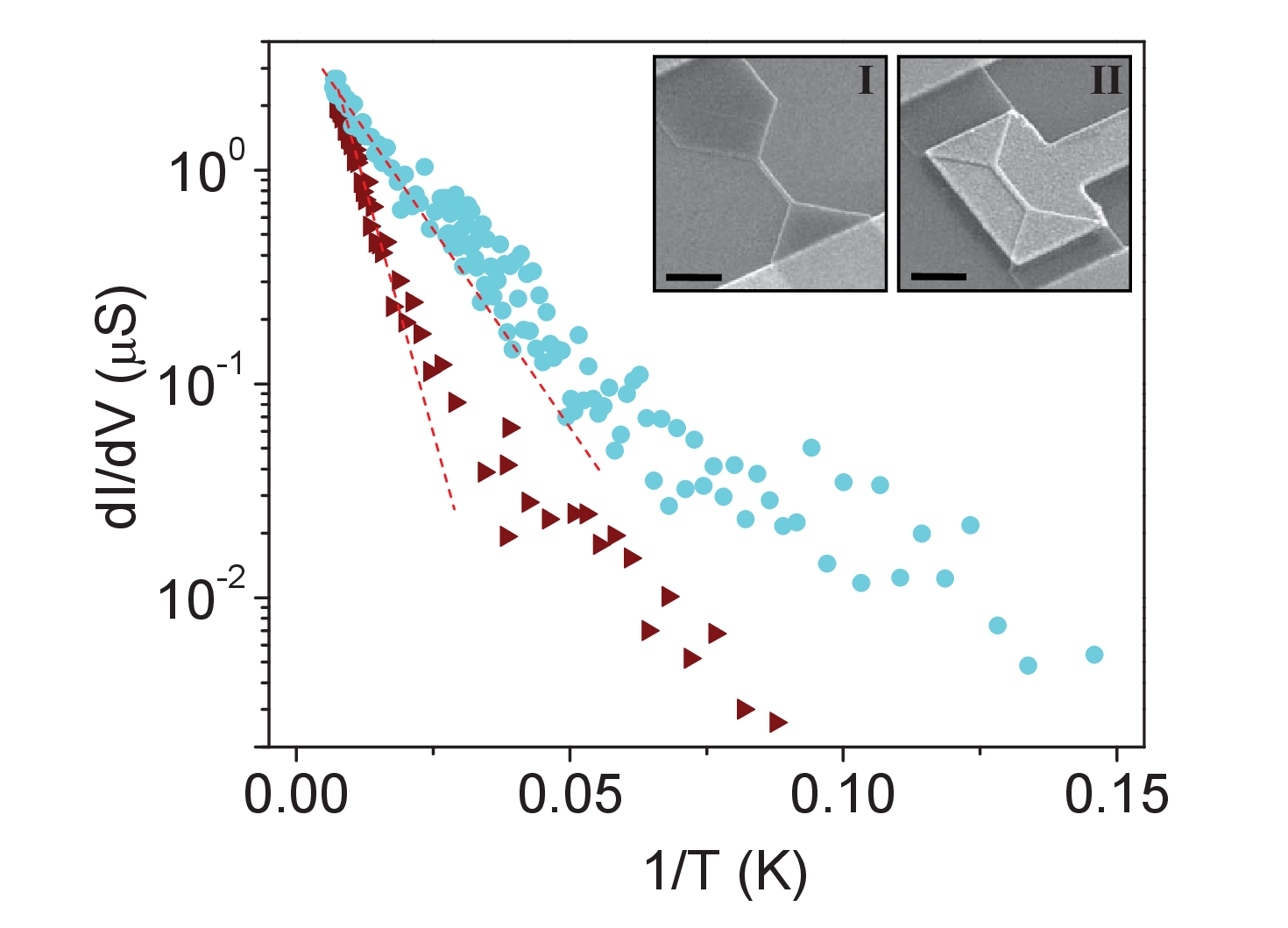}
\caption{ Temperature dependence of the conductance minimum for
dual gated (circles) and back gated (triangles) GNRs with the
similar $W$ and $L$. The dashed lines are Arrhenius fits in the
high temperature regime. The inset shows SEM images of back gated
(left) and dual gated (right) devices. Scale bar represents
500~nm.} \label{Fig.4}
\end{figure}

Finally, we discuss the effect of Coulomb charging in GNRs.
Several previous works have discussed the role of Coulomb blockade
and charging effects on the transport gap in GNRs and graphene
constrictions~\cite{Sols07, Stampfer09, Gallagher09}. In
principle, in a GNR with hopping between localized states, we
expect Coulomb interactions to open a soft Coulomb gap near the
Fermi surface, which can be incorporated into the total hopping
energy $E_a$ in addition to the single particle energy level
spacing $t'/w$, so that $E_a \approx t'/w+E_c$, where $E_c$ is the
Coulomb charging energy.~\cite{Martin09, Fogler2004vrh,
Efros-Shklovskii85Book-p409}. In order to quantify the
contribution of charging energy $E_c$ to the hopping energy $E_a$,
we perform a comparative transport measurement on GNRs with
different gate coupling. Fig.~4 shows the temperature dependent
minimum conductance $G_{min}(T)$ for a back gate only GNR (device
I) and a GNR with both top and back gates (device II) with the
similar $W$ and $L$. While device I has usual capacitive coupling
to the back gate, (i.e., $C^I\approx C_g$), $C^{II}$ for device II
is much closer to the top gate, leading to a larger capacitance:
$C^{II}/C^I\approx 4$. From the thermally activated Arrhenius
behaviors in the high temperature regime (dashed lines), we obtain
the activation energies of the two devices, $E_a^I=15$~meV and
$E_a^{II}=8.4$~meV averaged over two devices of type I and four of
type II. Considering the smaller charging energy contribution for
a top gated device, smaller values of the activation energy are
indeed expected, if Coulomb effects are appreciable in the GNR.

Employing the ratio $E_a^{I}/E_a^{II}\approx0.5$, we now can
estimate the charging energy contribution quantitatively.
Assuming that the single particle energy level spacing $t'/w$ is
similar for both GNRs due to their similar dimensions, we obtain
$E_a^{II}-E_c^{II} = E_a^{I}-E_c^{I} = t'/w$, where the charging
energy ratio of device I and II are given by
$E_c^{I}/E_c^{II}=C^{II}/C^{I}\approx 4$. The resulting estimate
for the charging energy contribution, $E_c^I/E_a^I\approx 0.6$,
indicates that the Coulomb charging effect provides a substantial
portion of the activation energy.

In conclusion, we investigate length and width dependent
resistance scaling in GNRs. Temperature dependent and electric
field dependent transport characteristics indicate that charge
transport in the transport gap of the disordered GNR is dominated
by localized states, where the Coulomb charging effects play an
important role.

The authors thank M. Fogler, I. Martin, K. Ensslin, D.
Goldhaber-Gordon, A. Young, P. Cadden-Zimansky, I. Aleiner, and B.
Altshuler for helpful discussion. This work is supported by the
ONR MURI, FENA, NRI, DARPA CERA. Sample preparation was supported
by the DOE (DE-FG02-05ER46215). JCB was supported by CNPq, Brazil.


\begin{thebibliography}{text}
\bibitem{footnote1}
Since the typical size of the transport gap in the GNRs in this
work is much larger than the thermal energy, we ignore thermally
excited transport via delocalized states outside of the gap.
\bibitem{footnote2}
This offset can either be ascribed to the critical length scale of
the edge localized states~[3, 16], to the critical percolation
length scale~[20], or to simply experimental inaccuracy of the
width determination due to over-etching underneath the etch
mask~[3].
\end{thebibliography}
\end{document}